\documentclass[5p,times]{elsarticle}

\usepackage{amssymb}
\usepackage{multirow}
\usepackage{flushend}
\usepackage{url}

\usepackage{natbib}
\setcitestyle{authoryear,open={(},close={)}} 

\journal{Future Generation Computer Systems}

\begin{document}
\sloppy

\begin{frontmatter}

\title{Container Resource Allocation versus Performance of Data-intensive Applications on Different Cloud Servers}

\author[First]{Qing Wang}
\ead{qw@clemson.edu}

\author[First]{Snigdhaswin Kar}
\ead{skar@clemson.edu}

\author[First]{Prabodh Mishra}
\ead{pmishra@clemson.edu}

\author[First]{Caleb Linduff}
\ead{clinduf@clemson.edu}

\author[First]{Ryan Izard}
\ead{rizard@clemson.edu}

\author[First]{Khayam Anjam}
\ead{kanjam@clemson.edu}

\author[First]{Geddings Barrineau}
\ead{cbarrin@clemson.edu}

\author[First]{Junaid Zulfiqar}
\ead{jzulfiq@clemson.edu }

\author[First]{Kuang-Ching Wang\corref{cor1}}
\ead{kwang@clemson.edu}
\cortext[cor1]{Correspondence to: Kuang-Ching Wang, 308 Fluor Daniel Building, Clemson University, Clemson, SC, 29634, USA}

\affiliation[First]{organization={Department of Electrical and Computer Engineering, Clemson University},
            city={Clemson},
            postcode={29634-0915}, 
            state={SC},
            country={USA}}


\begin{abstract}
In recent years, data-intensive applications have been increasingly deployed on cloud systems. Such applications utilize significant compute, memory, and I/O resources to process large volumes of data. Optimizing the performance and cost-efficiency for such applications is a non-trivial problem. The problem becomes even more challenging with the increasing use of containers, which are popular due to their lower operational overheads and faster boot speed at the cost of weaker resource assurances for the hosted applications. In this paper, two containerized data-intensive applications with very different performance objectives and resource needs were studied on cloud servers with Docker containers running on Intel Xeon E5 and AMD EPYC Rome multi-core processors with a range of CPU, memory, and I/O configurations. Primary findings from our experiments include: 1) Allocating multiple cores to a compute-intensive application can improve performance, but only if the cores do not contend for the same caches, and the optimal core counts depend on the specific workload; 2) allocating more memory to a memory-intensive application than its deterministic data workload does not further improve performance; however, 3) having multiple such memory-intensive containers on the same server can lead to cache and memory bus contention leading to significant and volatile performance degradation. The comparative observations on Intel and AMD servers provided insights into trade-offs between larger numbers of distributed chiplets interconnected with higher speed buses (AMD) and larger numbers of centrally integrated cores and caches with lesser speed buses (Intel). For the two types of applications studied, the more distributed caches and faster data buses have benefited the deployment of larger numbers of containers.
\end{abstract}


\end{frontmatter}

\section{Introduction}
Cloud computing has enabled the tremendous growth of cloud-based applications, achieving superb cost efficiencies and scalability thanks to virtualization technologies such as virtual machines (VM) and containers. Containers are increasingly more popular than VMs due to their faster start-up time, smaller sizes, and simpler life-cycle management. These benefits, however, come with a weaker resource isolation model where resource sharing amongst coexisting containers has only soft limits and assurances~\cite{morabito_hypervisors_2015, felter_updated_2015}. While the model presents a practical trade-off for many applications, potential impacts for the emerging class of "data-intensive" applications, e.g., artificial intelligence (AI), machine learning (ML), and big data transfer applications, have not been adequately studied. Such applications, by design, often consume as many resources as are feasible to handle the sheer volume of data processing and transfer over the network. When placing multiple containers on the same server with multiple CPU cores and a multitude of memory, storage, and networks, contention over these resources can lead to undesirable consequences if left uncontrolled.

Past research has studied container performance with respect to cloud server resources such as CPU, memory, disk~\cite{xavier2015performance, morabito2017virtualization, rausch2021optimized}, and I/O~\cite{monsalve_dynamic_2015, wang_toward_2018, china_venkanna_varma_analysis_2016}. Recently, studies also started to explore container performance in fog/edge resources such as GPUs, FPGAs, and TPUs for data analytics and deep learning algorithms \cite{varghese2018accelerator}. Efforts are also underway to develop better tools, such as cgroup~\cite{Linux_cgroup}, to improve containers' resource control and isolation. No studies to date, however, have assessed the resource contention problem for containerized data-intensive applications. We consider data-intensive applications that ingest, process, and sometimes send out large amounts of data. While different applications will surely differ in their data processing and movement patterns, similar contention patterns are expected. In this paper, we study two data-intensive applications with very different purposes - a deep-learning-based speech recognition application named Kaldi and a software-defined networking-based data transfer application named SOS. Through measurements of these applications on {\it CloudLab}, a U.S. National Science Foundation (NSF) future cloud testbed~\cite{duplyakin_design_2019}, this paper conducts systematic sweeps of configurable container attributes to measure each application's performance metrics, resource usage, and "under-the-hood" system events that can help explain their dependencies and dynamic patterns.

This paper is neither a bench-marking study with standardized workloads nor a study of resource allocation methods (e.g., CPU~\cite{monsalve_dynamic_2015}, memory~\cite{wang_toward_2018}, ~\cite{mao_resource_2016}) or container parameter configurations ~\cite{tao_dynamic_2017}.  Instead, this paper aims to give a close-up look at the consumption and contention of a multitude of resources in two typical yet significantly different, data-intensive applications. The focus is on the methodology for experimentation, measurement, and bottleneck diagnosis for such applications in different cloud systems. The paper alone would not provide a full conclusion of optimal deployment strategies for the vast range of diverse applications. Nonetheless, the methods and processes presented in this paper can be similarly applied to characterize any other data-intensive applications in any other cloud environments to derive the necessary insights for deployment guidance. Therefore, this paper's contributions include:

\begin{itemize}
  \item A survey of data-intensive application and container performance limits, providing adequate background and clarity for interpreting the experimental findings. 

  \item A synopsis of two data-intensive applications' architecture, data workflow, and their experimental topologies on CloudLab. 

  \item In-depth discussions of container resource contention from multiple measured metrics, highlighting the primary bottlenecks and performance-resource dependencies. 
  
  \item A complete process for characterizing other data-intensive applications for optimizing their deployment on diverse cloud systems.
\end{itemize} 

The rest of the paper is organized as follows. Section~\ref{sec:lit} presents the survey of data-intensive applications and container scaling studies in modern multi-core clouds. Section~\ref{sec:prob} formulates the problem from a cloud operator's perspective. Section~\ref{sec:exp} describes our experimental methodology and setups. Section~\ref{sec:results} presents and discusses the experiment results. Section~\ref{sec:intel-vs-amd} summarizes key differences observed between Intel and AMD processors. Section~\ref{sec:Dep} discusses the methodology for measuring data-intensive container applications and extracting useful clues for optimizing deployment. Section~\ref{sec:con} concludes the paper.

\section{Literature Survey}
\label{sec:lit}
This literature survey below conveys an up-to-date understanding of data-intensive application performances, container performance bottlenecks, and container orchestration mechanisms.

\subsection{Performance Bottlenecks for Data-intensive Applications}
Data-intensive applications incur significant data movement and data processing. The former calls for fast data transfer applications. GridFTP is one such application that has been widely employed in high-performance computing communities~\cite{allcock_globus_2005}. For the latter, machine learning algorithms utilizing multi-core CPUs, GPUs, and large memory are typical.

Due to its broad use, GridFTP has been extensively studied to optimize its data transfer performance. Such studies have examined methods to tune GridFTP parameters, typically on a dedicated bare-metal server referred to as a data transfer node (DTN), to maximize its end-to-end data transfer throughput~\cite{yildirim_end--end_2012, yildirim_how_2012, yildirim_application-level_2016}. These studies have focused on the performance fluctuations caused by external network status~\cite{nine_hysteresis-based_2015} with respect to GridFTP's pipelining, parallelism, and concurrency parameters. In~\cite{bresnahan_managed_2011}, a managed GridFTP service takes into account the CPU and memory usage status on data transfer clients to optimize the scheduling of GridFTP server resources. Essentially, GridFTP optimization has been seen as a scheduling problem on servers and clients with certain amounts of resources.

Machine learning algorithms expend significant time in moving data across storage/disks, memory/DRAM, and on-processor caches~\cite{povey_kaldi_2011,yazdani_unfold_2017}. Such applications process large data models (e.g., search graphs, neural networks) that are often larger than the on-chip cache (e.g., 256 MB for recent AMD Rome processors), incurring longer latency for DRAM access or, worse, disk-to-memory page swaps. Most data processing performance studies have focused on acceleration methods within the algorithms themselves~\cite{tang_deep_2019} or hardware accelerators~\cite{yazdani_ultra_2016}. On modern multi-core servers, inappropriate CPU affinity~\cite{saini_impact_2011} strategies could also cause cache contention and negatively impact performance. Past studies have resorted to scheduling processor cores for specific tasks to avoid cache contention among different tasks~\cite{brecht_evaluating_2006, hanford_characterizing_2013}.

\subsection{Performance Bottlenecks for Containers}
\label{sec:container-bottlenecks}
Containers support resource isolation primarily through namespace~\cite{Linux_Namespaces} and cgroups~\cite{Linux_cgroup}. The isolation model, however, is weaker than that of VMs. A namespace is used to create a container that has no visibility or access to objects outside itself, while cgroup is used to create groups of containers with a specified fraction of resources (i.e., memory, CPU, or I/O). With increasing numbers of containers running on the same CPU, studies have found a significant decrease in expected performance ~\cite{zhenyun_zhuang_taming_2017, china_venkanna_varma_analysis_2016}.

With cgroups, the following pitfalls have been identified: (1) cgroup memory allocation imposes an upper bound of memory usage without reserving memory for said containers; (2) cgroup manages anonymous memory and cached pages in a common memory pool without separate capping, and the former can evict the latter; (3) the operating system can reclaim memory from any cgroups when needed, regardless if that cgroups have reached their memory limit or not. As a result, all container cgroups still need to compete for memory with each other or with the operating system kernel. Unlike VMs, containers are not bounded by a virtualization abstraction when accessing resources. In ~\cite{huang_adaptive_2019}, it was pointed out that memory greedy applications can go beyond cgroup's memory limit and impact all containers on the same server.

The underlying kernel can also become an issue when scaling containers. Components such as page cache or buffer I/O are container agnostic, with no rate throttling mechanism in cgroup. If page cache is evicted due to memory pressure, or if one running container has overloaded I/O, all other containers in the system will be impacted. Recently, cgroup V2~\cite{Linux_cgroup_v2} has been published to provide finer grain resource controls. Similarly, the shared network stack in the kernel is found to be a major bottleneck ~\cite{hu_towards_2017}. As containers run on multi-core processors, parallel processing is essential. The lack of parallelization in the kernel network stack thereby limits the container network from scaling~\cite{lei_tackling_2019}.

\subsection{Container Orchestration Limitations}

Container orchestration addresses resource allocation for deployed containers as well as their auto-scaling and life cycle management. In addition to namespace and cgroups, cloud operators can also configure the same resources through APIs provided by the container platform (e.g., Docker, Kubernetes). As discussed in Section~\ref{sec:container-bottlenecks}, such allocation does not protect containers from contention in the cache, memory access latency, network bandwidth, or PCIe bus pressure.

\section{Problem Statement}
\label{sec:prob}

The first step towards a strategy for optimal management of containerized data-intensive applications in cloud systems is an accurate understanding of the nature of resource consumption and contention of such applications. This paper derives a close-up look at a multitude of resources by two typical yet significantly different, data-intensive applications. The measurement methods and lessons in this paper provide a blueprint for characterizing other data-intensive applications and cloud environments. The paper examines CPU, memory, and I/O resources on cloud servers based on, specifically, the latest Intel Xeon and AMD EPYC multi-core processors. The following questions are studied from a cloud operator's perspective:
\begin{itemize}
  \item {\bf For each containerized application, what are the key resources required, and how do they affect performance?}  Specifically, the relationships among properties of a data-intensive application's data flow, parallel processes, and consumed resources are studied. 

  \item {\bf For each resource, how can they be controlled when containers are deployed?} Specifically, a range of container deployment parameters, including placement options, are studied.

  \item {\bf For each server, how does its processor and system architecture affect container deployment decisions?} Specifically, the performance implication with respect to properties of processor cores, memory (including caches), and I/O buses (PCIe) are studied.
\end{itemize} 

Docker version 19.03.11 container engine is used in this study. While comparison with other container engines is beyond the scope of this paper, a discussion of Docker's container deployment workflow and control options is provided.

\section{Experimental Methodology}
\label{sec:exp}
The NSF CloudLab testbed~\cite{duplyakin_design_2019} is used to conduct experiments in this paper.  CloudLab is a cloud cluster with almost 1,000 machines distributed across three sites around the United States: Utah, Wisconsin, and South Carolina, employing state-of-the-art multi-core servers, storage, memory, and network elements from multiple vendors.  While the three CloudLab sites are connected by 100 Gbps networks, experiments in this study are conducted on, but not between, two sites (University of Utah, Utah; Clemson University, South Carolina).

\subsection{Servers and Processors}

The experiments are conducted as follows:
\begin{itemize}
\item SOS experiments are conducted with Intel Xeon processors (Dell c8220 servers) at the Clemson site;
\item Kaldi experiments are conducted with AMD EPYC Rome processors (Dell d6515 servers) at the Utah site;
\end{itemize}
The rationale of conducting SOS experiments at Clemson is due to the need of Dell's OpenFlow switches which are only available at the Clemson site. 

The c8220 server at Clemson consists of two Intel Xeon E5-2660 V2 multi-core processors, each with ten cores at 2.2 GHz. Its hardware configuration is shown in Table~\ref{CloudLab_Hardware_Intel} below. Docker and Kubernetes are installed on the nodes for our experiments. 
\begin{table}[htbp] 
    \caption{\label{CloudLab_Hardware_Intel}Configuration of CloudLab c8220 Node with Intel Xeon E5 multi-core processor}
    \centering
    \small
    \begin{tabular}[t]{|c|c|}
        \hline
        \textbf{\textit{Hardware}} & \multirow{2}{*}{\textbf{\textit{Specifications}}} \\
        \textbf{\textit{Types}} & \\
        \hline
            CPU cores & 2 Intel E5-2660 v2 10-core CPUs at 2.20 GHz \\
        \hline
            RAM& 256GB ECC Memory (16 X 16 DDR4) \\
        \hline
            Disk& Two 1 TB 7.2K RPM 3G SATA HDDs \\
        \hline
            NIC& Dual-port Intel 10Gbe NIC (PCIe v3.0, 8 lanes) \\ 
        \hline
    \end{tabular}
\end{table}

The d6515 server at Utah consists of an AMD EPYC Rome multi-core processor with 32 cores at 2.35 GHz. The simultaneous multi-threading technology is enabled by default. Its hardware configuration is shown in Table~\ref{CloudLab_Hardware_AMD} below. Docker and Kubernetes are installed on the nodes for our experiments. 

\begin{table}[htbp]  
    \caption{\label{CloudLab_Hardware_AMD}Configuration of CloudLab d6515 Node with AMD EYPC Rome multi-core Processor}
    \centering
    \small
    \begin{tabular}[t]{|c|c|}
        \hline
        \textbf{\textit{Hardware}} & \multirow{2}{*}{\textbf{\textit{Specifications}}} \\
        \textbf{\textit{Types}} & \\
        \hline
            CPU cores& 32-core AMD EPYC 7452 at 2.35GHz \\
        \hline
            RAM& 128GB ECC Memory (8 x DDR4 3200MT/s ) \\
        \hline
            Disk& Two 480 GB 6G SATA SSDs \\
        \hline
            NIC& Mellanox ConnectX-5 100 GB NIC (PCIe v4.0) \\ 
        \hline
    \end{tabular}
\end{table}

\subsection{Intel and AMD Processor Architecture}
\begin{figure}[htbp]   
\begin{center}
\includegraphics[width=1.0\columnwidth]{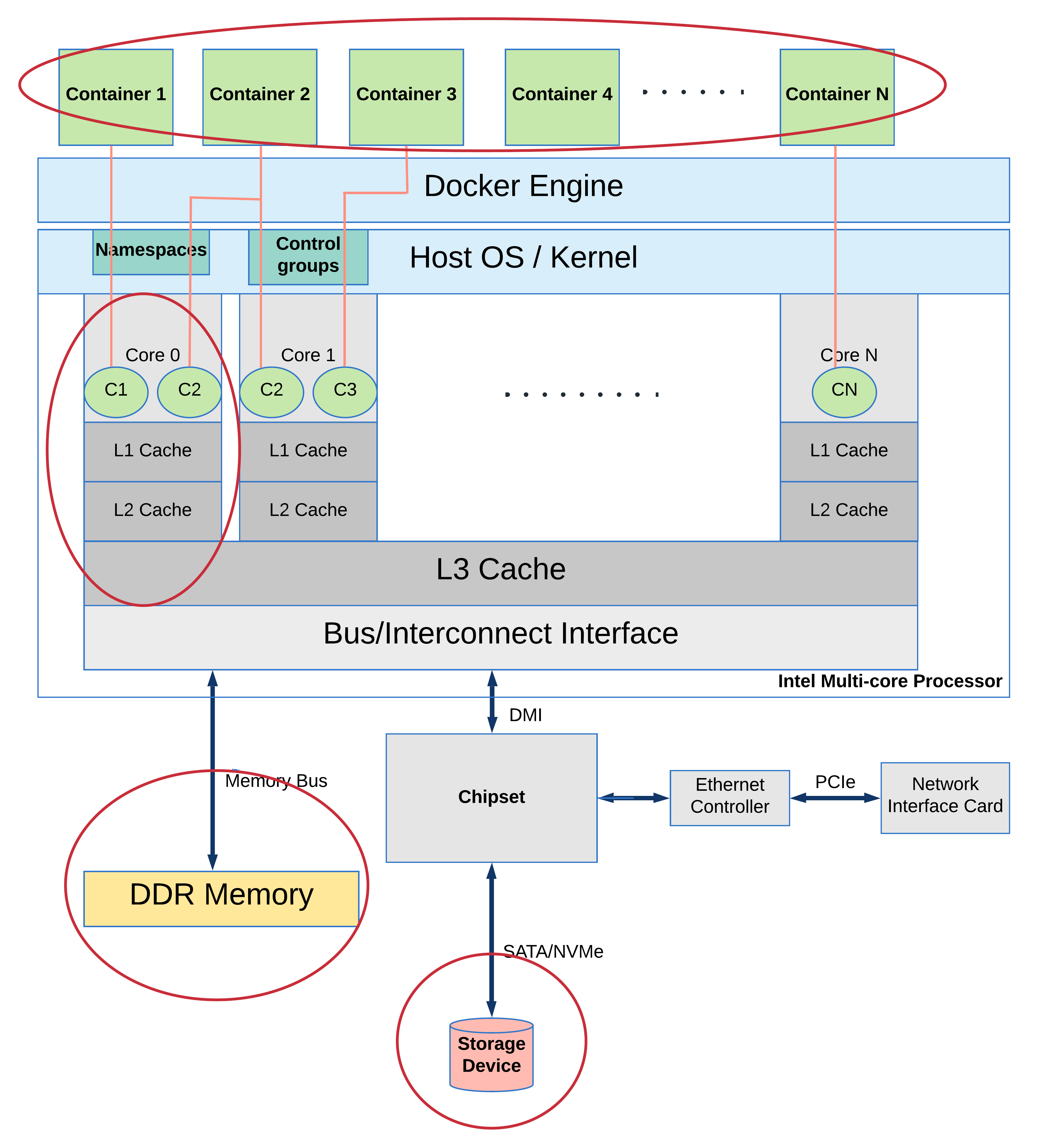}
\caption{Intel Xeon E5 multi-core system's relevant factors for containerized data transfer performance.}
\label{Intel_arch_image}
\end{center}
\end{figure}

Figure~\ref{Intel_arch_image} depicts the Intel Xeon E5 multi-core processor architecture. The network interface card (NIC) connects with the Intel chipset with a PCIe 3.0 bus. Hard disk drive (HDD) and solid-state drive (SDD) storage devices connect with the processor chipset over SATA/NVMe connections. SATA has up to 6 GB/s speed, while NVMe is 3.27~5.45 times faster than SATA. The chipset connects to the processor via a DMI 2.0 link, which has 2 GB/s. Hyper-threading~\cite{marr_hyper-threading_2002} is enabled by default in Intel E5, so each physical core presents as two logical cores in the system. Each logical core has its own fixed amount of L1/L2 cache, while all cores share a fixed amount of L3 cache and main memory. All cores can access the main memory and storage system (i.e., Linux swap space) via a shared system bus interface. Main memory (DDR3/DDR4) directly connects to Intel E5 via the memory bus. 

\begin{figure*}[t]
\begin{center}
\includegraphics[width=1.4\columnwidth]{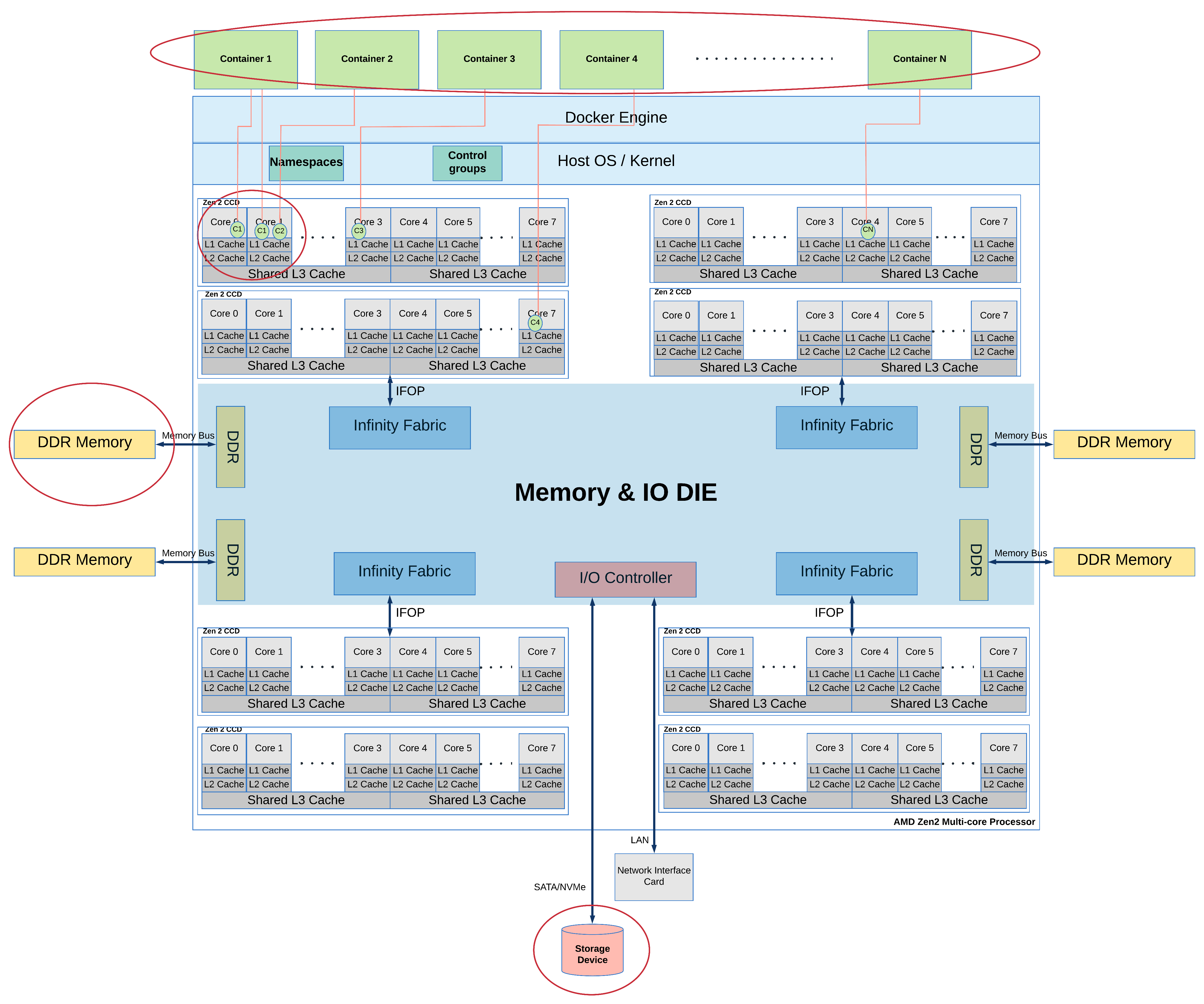}
\caption{AMD Zen2-based EPYC Rome multi-core system's relevant factors for containerized data transfer performance.}
\label{AMD_arch_image}
\end{center}
\end{figure*}

Figure~\ref{AMD_arch_image} depicts the AMD Zen2-based EPYC Rome multi-core processor architecture. AMD makes use of multiple chiplet dies to build up the whole processor. Each processor contains a centralized I/O die, which symmetrically pairs with up to eight CPU core die (CCD) chiplets. Each CCD and the I/O die connect via an independent Infinity Fabric On-Package (IFOP) link. The I/O die symmetrically provides eight DDR memory channels with the maximum access of 4TB of high-speed memory, as well as offers up to 128 PCIe 4.0 lanes of high-speed I/O. Because that single I/O die integrates all the I/O functionalities, no external chipset is needed for the processor chip. In the AMD EPYC processor, each CCD chiplet is composed of two CPU-Complexes (CCX). Each CCX can have up to four CPU cores, and each core has its own fixed amount of L1/L2 cache. Up to four cores in each CCX share up to 16MB L3 cache, and two CCXs inside each CCD chiplet have their separate L3 cache. The total L3 cache in an AMD Zen2-based multi-core processor can be 256 MB. Every CCD can have up to 8 cores, and up to 8 CCDs can be stacked onto a processor chip, so, in total, the CPU has up to 64 cores. Simultaneous multithreading (SMT) is enabled by default, so there can be up to 128 logical cores in the system. While all the CCDs, memory, and I/O connect to the single I/O die, they may be abstracted as separate quadrants, each with two CCDs, two memory channels, and 32 I/O lanes. All CCDs can connect to memory and I/O subsystem via the I/O die with the same average latency. 

\subsection{Intel and AMD Processor Comparison}
The Intel and AMD processors studied have fundamental differences. Our intention is not to compare their performance against each other. Instead, our focus is to highlight the different ways a containerized application would stress different resources on processors of different architectures. Therefore, accurate knowledge of the two processors' architecture remains critical.

Intel E5 has a monolithic architecture, while the AMD EPYC is composed of chip-lets. The AMD processor is more scalable and allows higher core count, larger memory bandwidth, and I/O for greater container density. The trade-off is that it may introduce additional inter-die memory access latency, which is mitigated by the use of a much larger L3 cache. For I/O, the Intel processor uses an external chip set, which may result in additional communication costs.

To date, Intel Xeon only supports PCIe 3, while AMD EPYC has supported 128 lanes of PCIe 4 support, doubling the I/O bandwidth from PCIe 3. 
AMD EPYC Rome has more PCIe lanes, so it can connect multiple NVMe/SATA devices without external chip support. AMD's centralized I/O die also features high-speed memory channels of DDR4-3200 with 204GB/s throughput, while Intel Xeon products can have around 120GB/s memory bandwidth - it gives AMD processor advantages in memory bandwidth-intensive workloads (i.e., deep neutral network modeling application). Finally, AMD enhances its Infinity Fabric bandwidth to meet the latency requirement as it interconnects all CPU dies and memories together with a single I/O die.

\subsection{Two Containerized Applications}
The two data-intensive applications are SOS (parallel data transfer) and Kaldi (deep neural network-based speech recognition).

\subsubsection{Parallel Data Transfer using Containers}
SOS (Steroid OpenFlow Service)~\cite{izard_steroid_2016} is an OpenFlow-based and agent-based service that transparently increases a TCP connection's throughput over large delay-bandwidth-product networks using parallel TCP connections. End-to-end TCP throughput is the primary performance measure for SOS.

\begin{figure}[htbp]
\begin{center}
\includegraphics[width=1.0\columnwidth]{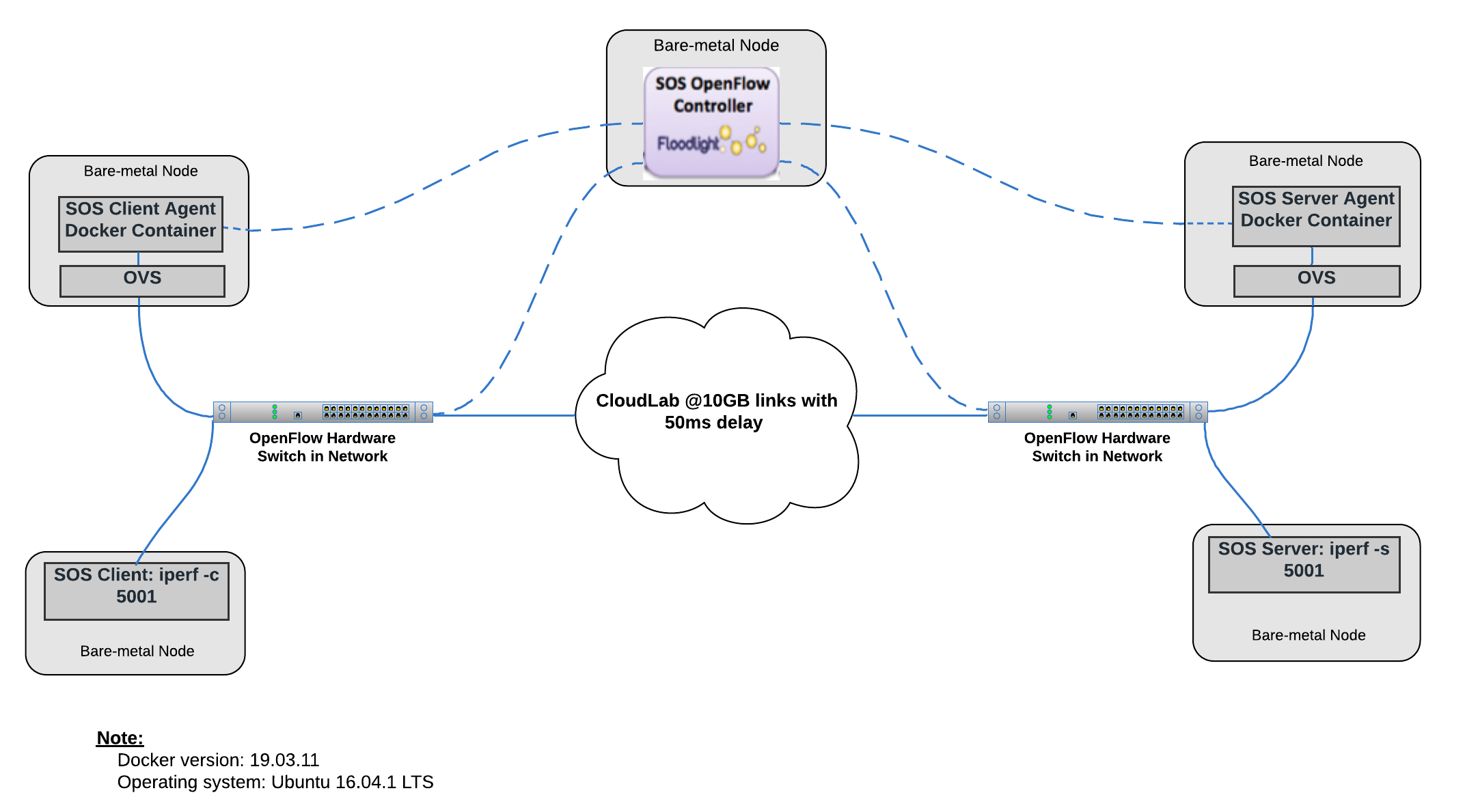}
\caption{Containerized SOS deployment topology in the CloudLab testbed}
\label{SOS_Topo}
\end{center}
\end{figure}

Figure~\ref{SOS_Topo} shows our containerized SOS deployment on CloudLab. It consists of a single Dell OpenFlow switch at each side of the network with directly connected bare-metal servers running the SOS agent application. TCP-based data transfer is initiated by a client/server application running on two bare-metal servers on the two ends. Open vSwitch (OVS)~\cite{OVS} version 2.6.0 is installed on each SOS agent node. All OpenFlow switches and OVSes connect to the SOS Floodlight~\cite{Floodlight} OpenFlow controller running on a separate bare-metal node. The Floodlight controller programs the switches to white-listed the permitted TCP flows. 50ms latency is simulated between the two Dell switches. Below is a summary of the steps taken to facilitate the SOS workflow:

\begin{itemize}
    \item \textbf{Step 1}: The Floodlight controller receives the request from a user via its whitelist API and installs all whitelisted TCP flow rules to OpenFlow switches in the network. The flow rules allow the OpenFlow switches to transparently convert the single TCP session between the client and server hosts into multiple parallel TCP sessions between the two containerized SOS agents.
    
    \item \textbf{Step 2}: The client and server hosts begin the TCP data transfer. OpenFlow switches rewrite each TCP packet's header.
    
    \item \textbf{Step 3}: The SOS agent containers ramp up the parallel TCP sessions' throughput to maximize the use of the available link bandwidth. 
\end{itemize}

In step 3, the SOS agent containers must perform per-packet rewrite operations. On the sending side, the client or server application retrieves data from the storage (HDD or SDD) and sends the data to the network over the NIC. The SOS agent container then performs the per-packet rewrites, incurring CPU/kernel overheads. Hence, the underlying CPU, NIC, and the memory data bus would affect the SOS data transfer performance. On a multi-core cloud server, the agent containers can distribute their computation task across server CPU cores. This can create the memory bus (i.e., PCIe bus) pressure between the network card and the CPU. 

The OS kernel tasks and other system root processes may be interrupted or compete with SOS container processes, slowing down the SOS packet processing speed performance. Finally, an inappropriate CPU allocation may lead to cache contention. 

\subsubsection{Distributed Data Process using Containers}
Kaldi is a speech recognition toolkit.
Total speech-to-text task completion time is the primary performance measure for Kaldi.

Figure~\ref{Kaldi_Topo} shows our containerized Kaldi deployment on CloudLab. It is a typical Kaldi deployment that consists of a single master node and four worker nodes. One Kaldi master container runs on a Dell server (the master node) with multiple audio files inside. Each audio file is 1080 seconds (18 minutes) long and is divided into 72 15-second chunks. When a user requests those audio files for speech conversion, the audio chunks are evenly distributed from the master node to four worker nodes via the scp utility so that they can be processed in parallel. Once the data chunks arrive at a worker node, Kaldi containers on that node will process the audio chunks and generate text outputs. Specifically, each container always handles 18 audio chunks. Based on the number of audio chunks received, each worker node would launch the needed number of Kaldi containers. Once each container completes its task, its text output is sent back to the master node. The master container will combine all outputs to produce a complete transcript of the audio file. Below is a summary of the steps taken by the Kaldi speech recognition service:
\begin{figure}[htbp]
\begin{center}
\includegraphics[width=0.95\columnwidth]{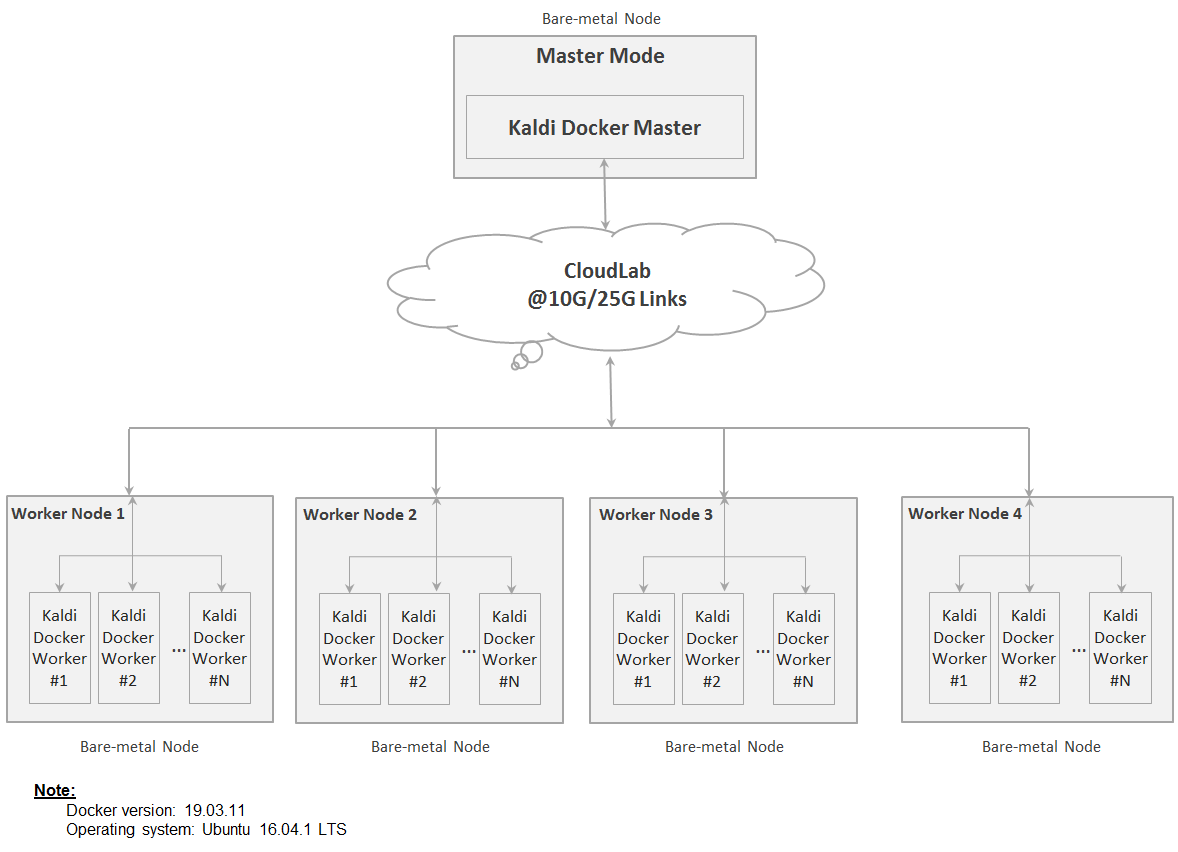}
\caption{Containerized Kaldi deployment topology in the CloudLab testbed}
\label{Kaldi_Topo}
\end{center}
\end{figure}

\begin{itemize} 
    \item \textbf{Step 1}: The master container transfers the audio files in chunks to a specified number of worker nodes via scp and sends the docker commands to the worker nodes to boot up the needed number of worker containers.  
    
    \item \textbf{Step 2}: A worker node receives the audio chunks and commands to launch the Kaldi containers. Each Kaldi container runs audio chunks through the speech recognition application and produces an output.
    
    \item \textbf{Step 3}: Each container cleans up its output of any Kaldi artifacts and sends it back to the master container.
    
    \item \textbf{Step 4}: The master container receives the outputs from all worker containers and pieces them together into the final text output. 
\end{itemize}

In particular, in step 3, Kaldi containers consume a large amount of CPU and memory resources on the worker nodes. By design, Kaldi speech recognition has three parts: (1) convert original speech to a phoneme representation; (2) convert phoneme representation to words after generating a word lattice. (3) a decoder processes the lattice to produce the most probable text output. (1) and (2) do not consume significant compute resources once the lattice is generated. In our setup, the Kaldi decoder runs in a per-chunk mode, meaning it loads a lattice when processing each incoming audio chunk. The size of a lattice is usually in gigabytes; each container consumes CPU and memory resources to buffer the model, then computes the text probability based on it. 
When scaling up Kaldi, increasing containers will experience bus bandwidth issues between CPU and other subsystems. The OS kernel tasks or other system root processes will also be interrupted, resulting in increased system context switches and performance overheads.

\subsection{Experimental Setup} 
The experiments are conducted on two CloudLab testbed sites with, respectively, Intel Xeon based and AMD EPYC Rome based servers. 

\subsubsection{Container CPU Allocation}
The SOS application experiments are done on Intel servers on CloudLab at Clemson. With an OpenFlow-based network connecting the servers, data transfer throughput is the primary performance metric of focus. 

Figure~\ref{SOS_Topo} shows the experiment topology. Five bare-metal servers are allocated for running, respectively, the SOS client, the SOS server, the SDN controller, the client-side SOS agent, and the server-side SOS agent. Each node has one Intel multi-core processor. A 10 Gbps link between the two SOS agents is configured to model a wide area network connection with 50 ms latency between the two SOS agents. All network interfaces along the five nodes are configured to have a maximum transmission unit (MTU) of 9000 bytes (a jumbo frame). The SOS agents are containerized. By assigning more CPU cores to an SOS agent container or deploying multiple pairs of SOS agents, higher data transfer throughput can be achieved. 

For SOS agents, the most intense operation is the per-packet copying (a system call) of received packets from the NIC to the CPU cores and vice versa. This can result in varying levels of cache contention, which varies with the CPU cores being used. With Intel hyperthreading, each thread is represented as a virtual core (vCPU), every two consecutive virtual CPUs (i.e., (0,1), (2,3), (4,5), etc.) share the same L1/L2 cache, and all vCPUs inside a processor socket share the same last level L3 cache. To study the SOS performance dependency on CPU allocation, we design the following sequence of experiments with different numbers and placements of CPU cores:
\begin{itemize}
    \item \textbf{Configuration 1}: Using the Docker default scheduler for vCPU core allocation. In this case, the SOS container computation task may use any available cores. 
    
    \item \textbf{Configuration 2}: Allocating only one vCPU core to the SOS agent container. In this case, the SOS container computation task is dispatched as one single hyper thread to that assigned CPU core.
    
    \item \textbf{Configuration 3}: Allocating two consecutive vCPU cores (i.e., core 0, 1 or core 2, 3) to the SOS agent container. In this case, the SOS container computation task is dispatched as two concurrent hyperthreads with the same L1/L2 cache. 
    
    \item \textbf{Configuration 4}: Allocating two non-consecutive vCPU cores (i.e., core 5 and core 15) in the same CPU socket to the SOS agent container. In this case, the SOS agent computation tasks on the two cores are dispatched to separate L1/L2 caches but the same last level L3 cache. 
    
    \item \textbf{Configuration 5}: Allocating two non-consecutive vCPU cores (i.e., core 10 and core 30) in different CPU sockets to the SOS agent container. In this case, the SOS agent computation task is totally dispatched to separate L1/L2/L3 caches. 
    
     \item \textbf{Configuration 6}: Allocating three vCPU cores (i.e., cores 2, 4, 8, or 2, 4, 28) to the SOS agent container. In this case, the SOS agent computation task is dispatched to separate L1/L2 caches, while the last level L3 cache may or may not be the same. 

     \item \textbf{Configuration 7}: Allocating four vCPU cores (i.e., cores 2, 4, 6, 8, or cores 2, 4, 6, 28) to the SOS agent container. In this case, the SOS agent computation task is dispatched to separate L1/L2 caches, while the last level L3 cache may or may not be the same.

     \item \textbf{Configuration 8}: Allocating five vCPU cores (i.e., cores 2, 4, 6, 8, 10, or cores 2, 4, 6, 28, 38) to the SOS agent container. In this case, the SOS agent computation task is dispatched to separate L1/L2 caches, while the last level L3 cache may or may not be the same. 
\end{itemize}

The {\it iperf} utility is used to generate TCP traffic for SOS container experiments. For each experiment, iperf runs for a 50 second duration, repeating five times for statistical confidence. 

\subsubsection{Container Memory Limits}
The Kaldi application experiments are done on AMD servers on CloudLab at the University of Utah as well as Intel servers at Clemson. The speech recognition completion time is the primary performance metric of focus. 

Figure~\ref{Kaldi_Topo} shows the experiment topology. Five bare-metal servers are allocated for running one containerized Kaldi master and four containerized Kaldi workers connected over a 25 Gbps network.

For Kaldi workers, the Kaldi decoder needs to load a very large-sized (i.e., around 1.3 GB in our setup) pre-trained model to perform the speech recognition task. The workers may experience out-of-memory (OOM) exceptions and significantly degrade their performance. To study the Kaldi performance dependency on memory settings, we vary the container memory limit set for each Kaldi worker container. 

The Kaldi application is both compute and memory intensive. Once the appropriate memory limit is confirmed from the previous experiment, the subsequent experiment measures rum-time system-wide metrics for CPU/memory/disk resources using nmon~\cite{noauthor_nmon_nodate} with both AMD and Intel servers. 

The experiment is conducted with the same five server topologies. While there remains one Kaldi master container, the number of Kaldi worker containers are varied on each worker server. The time series of measured run-time metrics are studied to understand the resource contention patterns for this class of applications. 

For each Kaldi container experiment, a script on the master is executed to split big audio files into smaller chunks. {\it scp} is then used to distribute the data chunks to respective worker nodes for processing. 

\subsection{Performance Metrics}

\begin{table}[htbp]
\caption{Container Performance Metrics And System Resource Statistics}
\begin{center}
\small
\begin{tabular}{|c|c|}
    \hline
    \textbf{\textit{Performance Metrics}} & \textbf{\textit{System Resources Statistics}} \\
    \hline
    \multirow{2}{*}{Transfer throughput (Gbps)} & CPU usage\\
                                                & Context switching\\
                                                & Page copy count\\
    \hline
    \multirow{4}{*}{Completion time (sec)}   & Page copy operations frequency\\
                                             & CPU usage\\
                                             & Memory usage\\
                                             & Page cache utilization\\
                                             & Disk busy rate\\
    \hline    
\end{tabular}
\end{center}
\label{Data_Collection}
\end{table}

The performance metrics for SOS and Kaldi experiments are listed in Table~\ref{Data_Collection}. 
For SOS, (1) system context switches and (2) system-wide page copy counts are monitored and recorded. The former is the indicator of system performance overhead. 
When the memory subsystem hits the bottleneck, the system-wide page copy count will be flattened out even if more CPU cores are allocated to the container.

\section{Experimental Results And Discussion}
\label{sec:results}
In this section, we examine each container experiment to derive important observations and explanations of container performance and scalability.

\begin{table}[htbp]
\begin{center}
\scriptsize
\caption{Containerized SOS performance of different CPU allocation strategies in Intel Xeon platform}
\begin{tabular}{|l|c|c|}
\hline
\multicolumn{1}{|c|}{\multirow{3}{*}{\textbf{\textit{CPU Allocation Strategies}}}} & \textbf{\textit{Avg.}} & \textbf{\textit{System }} \\
															 & \textbf{\textit{Throughput }} & \textbf{\textit{Context}} \\
															 & \textbf{\textit{(Gbps)}} & \textbf{\textit{Switching}} \\
\hline
\textbf{Configuration 1}: Docker default CPU scheduler may & \multirow{2}{*}{5.6} & \multirow{2}{*}{10320} \\
use any vCPU core on the node for a SOS container & & \\
\hline
\textbf{Configuration 2}: Allocate a single vCPU core with a & \multirow{2}{*}{5.47} & \multirow{2}{*}{6440} \\
single SOS container & & \\
\hline
\textbf{Configuration 3}: Allocate two consecutive vCPU cores & \multirow{3}{*}{3.49} & \multirow{3}{*}{66469} \\
for a SOS container, while select both cores from the & & \\
same L1/L2 cache & & \\
\hline
\textbf{Configuration 4}: Allocate two far-away vCPU cores for & \multirow{4}{*}{7.12} & \multirow{4}{*}{6957} \\
a SOS container, while select both cores from separate & & \\
L1/L2 cache but they reside in the same CPU socket & & \\
\hline
\textbf{Configuration 5}: Allocate two far-away vCPU cores for & \multirow{3}{*}{7.32} & \multirow{3}{*}{8441} \\
a SOS container, while these cores have separate & & \\
L1/L2 cache and reside in separate CPU sockets & & \\
\hline
\textbf{Configuration 6}: Allocate three vCPU cores for a SOS & \multirow{3}{*}{8.91} & \multirow{3}{*}{9546} \\
container, while these cores have separate L1/L2 & & \\
\hline
\textbf{Configuration 7}: Allocate four vCPU cores for a SOS & \multirow{3}{*}{8.76} & \multirow{3}{*}{10041} \\
container, while these cores have separate L1/L2 & & \\
\hline
\textbf{Configuration 8}: Allocate five vCPU cores for a SOS & \multirow{3}{*}{8.87} & \multirow{3}{*}{10663} \\
container, while these cores have separate L1/L2 & & \\
\hline
\end{tabular}
\label{SOS_CPU_Allocation_Intel}
\end{center}
\end{table}

\subsection{Container CPU Allocation Strategies} 
Table~\ref{SOS_CPU_Allocation_Intel} shows the measurement results of different CPU allocation strategies for SOS running on Intel servers. The default Docker CPU scheduler gives an average of 5.6 Gbps data transfer throughput, while an average of 5.47 Gbps throughput is observed when allocating a single vCPU core to the container. Subsequently, a performance degradation (down to 3.49 Gbps) was seen when allocating two adjacent vCPU cores (i.e., vCPU cores 2 and 3 that share the same L1/L2 cache) to that container. When allocating two far-away vCPU cores to an SOS container, we observe a 7.12 Gbps performance if those two cores (i.e., vCPU cores 5 and 15) are located on separated L1/L2 cache while within the same CPU processor socket, and a 7.32 Gbps performance if those two cores (i.e., vCPU core 10 and 30) are located on separate L1/L2 cache from separated sockets. We observed 8.91 Gbps, 8.76 Gbps, and 8.87 Gbps network throughput performance when allocating three cores, four cores, and five cores, correspondingly. 

One interesting observation from table~\ref{SOS_CPU_Allocation_Intel} is that configuration 3 gives the worst performance: its performance is only 63\% compared with configuration 1 and is only 49\% compared with configuration 4. This phenomenon is likely due to cache level contention. This is further confirmed by the abrupt increase in context switching, which indicates a large system kernel overhead due to the cache level competition.

The results overall confirms the significance of CPU allocation configuration for such applications. Pinning more vCPU cores to an SOS agent container has clear and significant performance benefits.
Allocating more than three CPU cores did not yield a significant advantage. The flattening performance is likely due to the memory bus bandwidth becoming the bottleneck, saturated by the data copies incurred by the SOS agent.

\begin{figure}[htbp]
\begin{center}
\includegraphics[width=0.85\columnwidth]{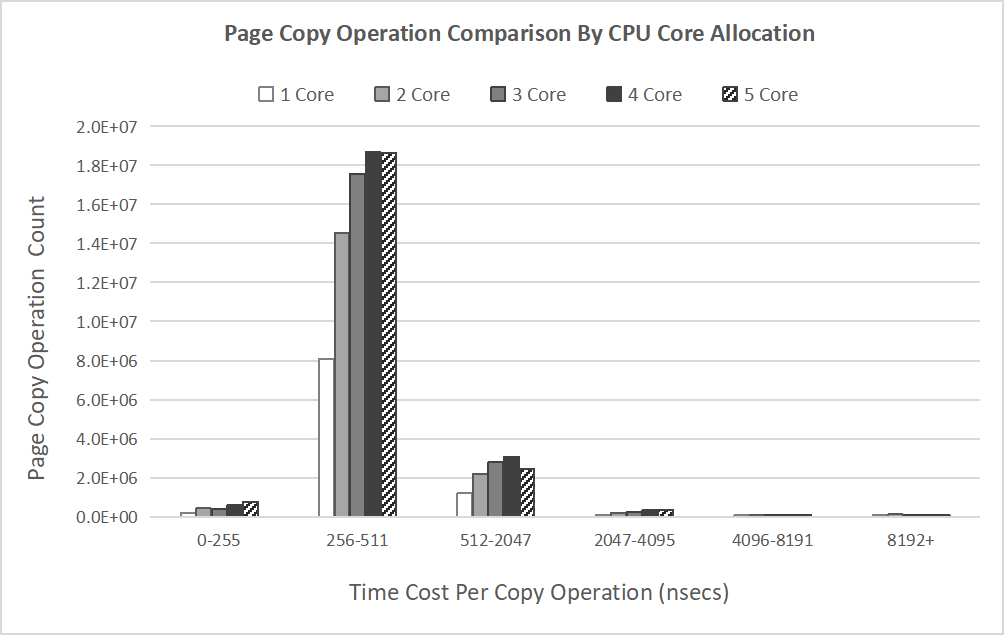}
\caption{System-wide Page Copy Count By Different CPU Core Allocation. Page counts are observed firstly observed increase (1.2 times more) when allocating one, two and three cores to the agent container. Page counts are flatten out when allocating more then three cores to the agent container.}
\label{SOS_PageCopyDist}
\end{center}
\end{figure}

Figure~\ref{SOS_PageCopyDist} shows system-wide page count statistics from five CPU allocation configurations. The 2-core allocation has two times the page copy operations versus single core allocation, while three core allocation has 1.2 times more page copies than the 2-core allocation case. The page copy numbers flatten out when allocating more than three cores, consistent with our performance observation in Table~\ref{SOS_CPU_Allocation_Intel}. 

In summary, for an SOS-like compute-intensive containerized application, the performance would benefit from distributing the computation task to multiple cores. However, cloud operators should make the CPU resource scheduling decision in a cache-aware manner: to isolate those containers based on the L1/L2 cache in order to reduce the possibility of cache contentions. To fully utilize the modern multi-core architecture, cloud operators should also be aware that the memory bus bandwidth is likely to be a throttling bottleneck.

\subsection{Container Memory Limit Control} 
\begin{figure}[t].
\begin{center}
\includegraphics[width=0.8\columnwidth]{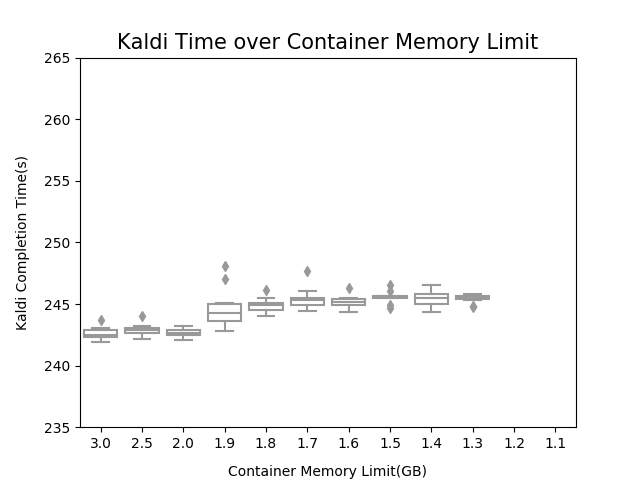}
\caption{Average Kaldi container performance (box and whisker plots) vs. container memory limit setting in Intel platform. The average Kaldi speech recognition time is almost constant but starts to experience an OOM when container memory limit is set to less than 1.3 GB.}
\label{ContainerPerformance_over_MemoryLimit_Intel}
\end{center}
\end{figure}

\begin{figure}[t].
\begin{center}
\includegraphics[width=0.8\columnwidth]{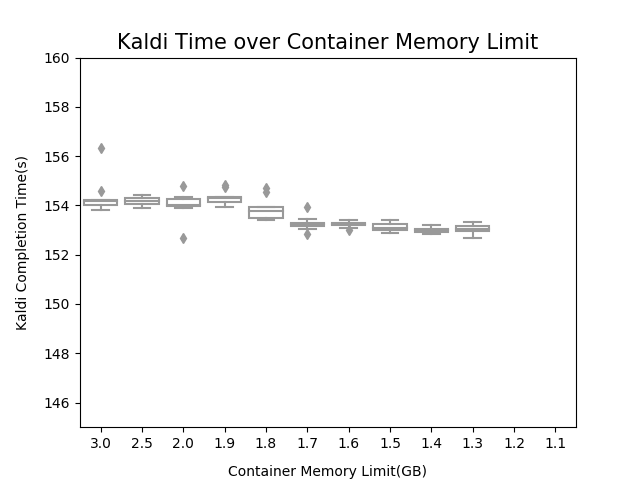}
\caption{Average Kaldi container performance (box and whisker plots) vs. container memory limit setting in AMD platform. The average Kaldi speech recognition time is almost constant but starts to experience an OOM when container memory limit is set to less than 1.3 GB.}
\label{ContainerPerformance_over_MemoryLimit_AMD}
\end{center}
\end{figure}

Figure~\ref{ContainerPerformance_over_MemoryLimit_Intel} shows a single Kaldi container performance on the Intel servers with different container memory limits from 3.0 GB to 1.1 GB. Kaldi container showed almost constant performance (around 255 seconds) to finish its job over different memory limit settings. However, it starts to experience an OOM error when the memory limit is less than 1.3 GB. This is because the Kaldi container does not have enough memory to load the pre-trained language models.   

Similarly, on AMD servers, a single Kaldi container with memory limits from 3.0 GB to 1.1 GB showed constant performance, as shown in Figure~\ref{ContainerPerformance_over_MemoryLimit_AMD}. Consistently, Kaldi starts to experience an OOM exception when the memory limits are set below 1.3 GB.  

In summary, a containerized Kaldi application performance is not sensitive to the container memory limits setting, as long as the setting satisfies a minimum memory specific to the application.

\subsection{Multiple Containers on an Intel Server}  
In this section, we examine the performance and resource statistics when a different number of Kaldi containers are deployed on an Intel server.

\subsubsection{Observations}
\begin{figure}[htbp].
\begin{center}
\includegraphics[width=0.95\columnwidth]{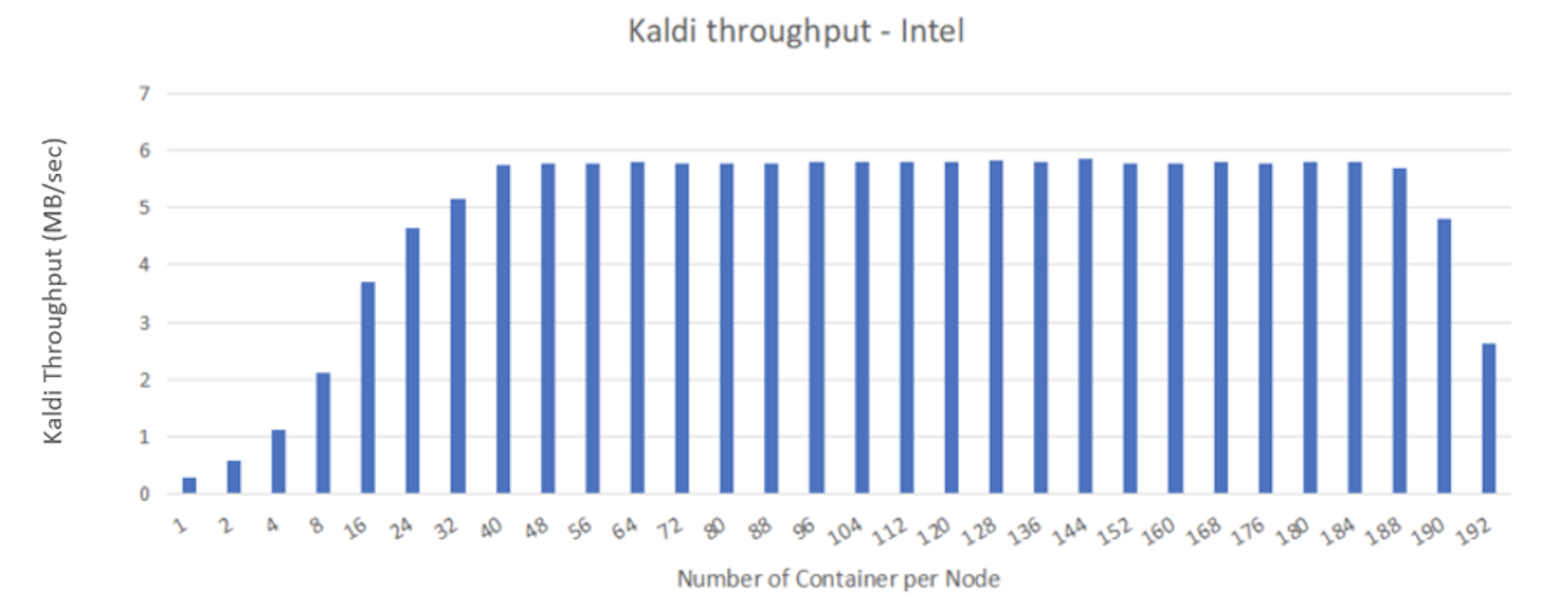}
\caption{Kaldi container performance vs. number of containers at Intel platform. Three levels of speech recognition completion time are observed: 1-to-40 containers as level 1, 40-to-188 containers as level 2, and 188-to-192 containers as level 3.}
\label{ContainerNumber_VS_KaldiPerformance_Intel}
\end{center}
\end{figure}

Figure~\ref{ContainerNumber_VS_KaldiPerformance_Intel} shows the Kaldi performance for different numbers of containers on a server. Each container memory allocation is set to 1.3 GB. Three levels of operation were observed: from 0 to 40 containers, the system throughput increases (0.5 MB/sec to 5.8 MB/sec). Then, from 40 to 188 containers, the system throughput flattens at around 5.8 MB/sec. Lastly, from 188 to 192 containers, the system throughput quickly diminished (5.8 MB/sec to 2.74 MB/sec) with increasing numbers of containers on the server.

\subsubsection{System Analysis}
\label{sec-page-cache-intel}
For more insights into the three levels of operation, Figure~\ref{ContainerNumber_VS_KaldiPerformance_Intel} shows the collected time series of a variety of system resource usage on the Intel server. 

\begin{figure*}[t].
\begin{center}
\includegraphics[width=1.4\columnwidth]{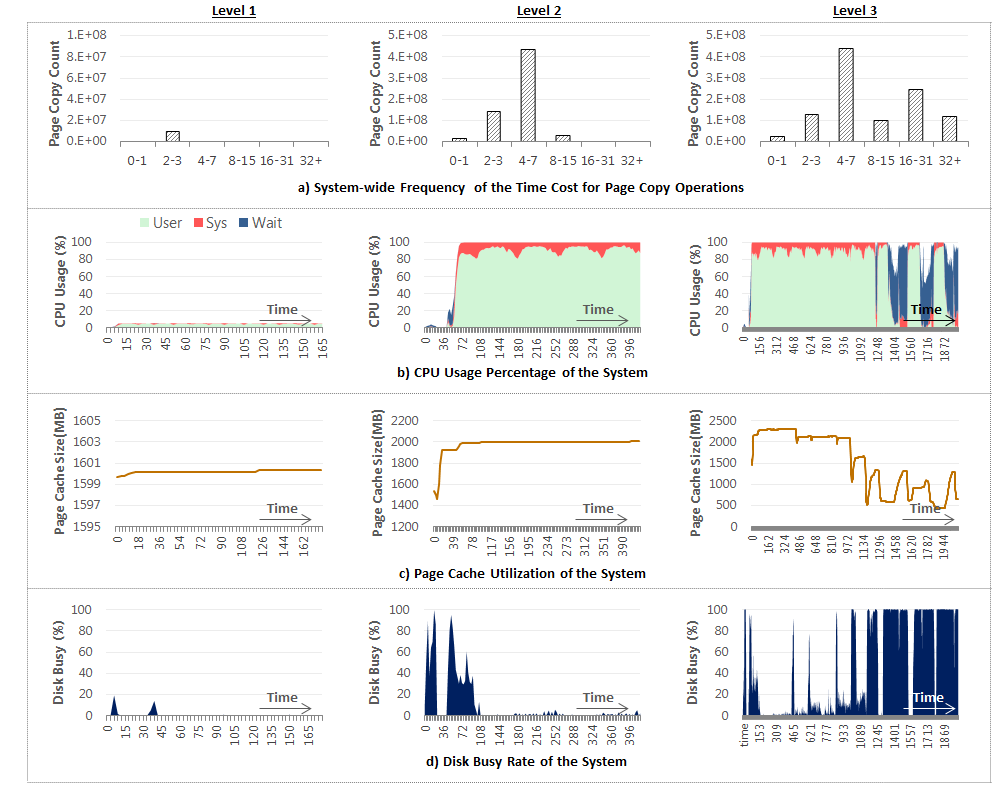}
\caption{System analysis plots with Kaldi containers on an Intel server: (a) the frequency of the time cost for page copy operations. (b) the CPU usage of the system (c) the page cache utilization over the time. (d) the disk busy rate of the system over the time.}
\label{ContaineScaling_System_Analysis_Intel}
\end{center}
\end{figure*}

Figure~\ref{ContaineScaling_System_Analysis_Intel}(a) summarizes the frequency of the time cost for each page copy operation of the whole system. The x-axis represents the time cost per page copy operation in a microseconds range (i.e., 0-1 us), and the y-axis represents the total number of page copy operations in that range. During speech recognition, each Kaldi container periodically loads the pre-trained models. Such pre-trained models are buffered in the virtual memory (``page cache" in Linux kernel) from the second time of access. So Kaldi containers can refer to the data directly from the page cache without disk access taking place. However, all containers in the system share the same page cache, which leads to intensive data copy operations from the kernel space to the user space. Those operational costs cannot be neglected, especially when the system is under a high CPU workload. With 2-container deployment, each page copy operation spends around 2 or 3 microseconds. The time cost pattern is significantly different for the 112 containers case, in which case most copy operations take 4 to 7 microseconds to finish, and the system CPU usage is already saturated. At 190 containers, a noticeable amount of copy operations takes 8 to 15 microseconds or more to finish. At level 1, the system is not fully loaded as not too many containers are deployed. Starting from level 2, the underlying system bus bandwidth is stressed. At level 3, the whole worker node system is fully stressed, which leads to an even worse bus bandwidth pressure. 

Figure~\ref{ContaineScaling_System_Analysis_Intel}(b) shows the CPU utilization. The utilization steadily grows in level 1, maxes out in level 2, and enters a wait state significantly in level 3 operation. 

Figure~\ref{ContaineScaling_System_Analysis_Intel}(c) and~\ref{ContaineScaling_System_Analysis_Intel}(d) show the system-wide page cache size and the disk busy rate over time. At levels 1 and 2, we observe a steady and high page cache utilization, combined with a low disk busy rate, indicating that all the deployed containers leverage the page cache well to load language models from RAM without disk access. The reason we have a burst at the beginning of level 2 of Figure~\ref{ContaineScaling_System_Analysis_Intel}(d) is because all the 112 containers have to read the language model from the disk for the first time; later, the model can be stored in the page cache buffer. At level 3, the page cache utilization changes dramatically over time, as Figure~\ref{ContaineScaling_System_Analysis_Intel}(c) shows. This significant page cache fluctuation starts from 188 Kaldi containers running on the server. If the page cache drops, consequently, all running containers have to interact with slow disk I/O, leading to a big performance penalty (i.e., 1390 seconds average completion time for 112 containers vs. 2851 seconds for 190 containers). 

The significant page cache fluctuation is also caused by the Linux cache eviction mechanism, which is triggered when the system is under severe memory pressure. Each Kaldi container memory limit allocation is 1.3 GB; however, such a memory limit setting only imposes the upper boundary of the container memory usage but not the memory reservation. At the container runtime later, the operating system allocates the container's requested memory on the fly. If too many containers are running concurrently on a server (i.e., 190 containers), and if those container applications are memory intensive, the operating system (e.g., 256GB ECC memory in Intel c8220 node) is severely stressed without enough free memory to allocate further requests. The operating system, therefore, has to free memory from all possible places, including the page cache buffer. Once page cache eviction happens, this reclamation leads to a system-wide impact for all the other running containers and will force them to interact with the slow disk I/O to retrieve the evicted items (language models). This is certainly not desired for scaling containers and leads to the Kaldi container slowing down. Moreover, all running Kaldi containers share the same kernel buffer for the disk I/O without isolation or rate throttling mechanism. When too many containers interact with disk I/O, the contention can happen in the buffer and result in performance overheads. Or, if a container/cgroup has a high disk I/O, that would interfere with all other containers running on the same server, making an even worse system impact for all other containers requiring disk I/O access. As a result, Kaldi container performance at level 3 will be severely degraded, as the tail part of Figure~\ref{ContainerNumber_VS_KaldiPerformance_Intel} shows. 

\subsubsection{Summary}
When scaling Kaldi containers on an Intel server, our results expose the non-negligible data copy operation overhead and the possible page cache eviction impacts. Analysis of the utilization of CPU, page cache usage, and disk busy rate reveals that the copy operation overhead is caused by the underlying bus bandwidth pressure between the CPU and subsystems such as memory and I/O. CPU utilization can be overloaded when deploying more than 190 containers in the Intel server, which results in busy waiting. A dedicated CPU core for OS kernel tasks and other system processes should mitigate this issue. Although page cache is a common Linux kernel component to enhance the data access efficiency, the page cache eviction can be triggered under a system memory pressure case and will force all influenced running containers to directly interact with the disk. A secondary disk I/O contention can happen as all the running containers share the same kernel buffer for disk I/O. From our observations, page cache eviction can negatively impact all the containers running on a cloud server, resulting in diminishing container performance.

\subsection{Multiple Containers on an AMD Server}  
In this section, we examine the performance and resource statistics when a different number of Kaldi containers are deployed on an AMD server.

\subsubsection{Observations} 
\begin{figure}[htbp]. 
\begin{center}
\includegraphics[width=0.95\columnwidth]{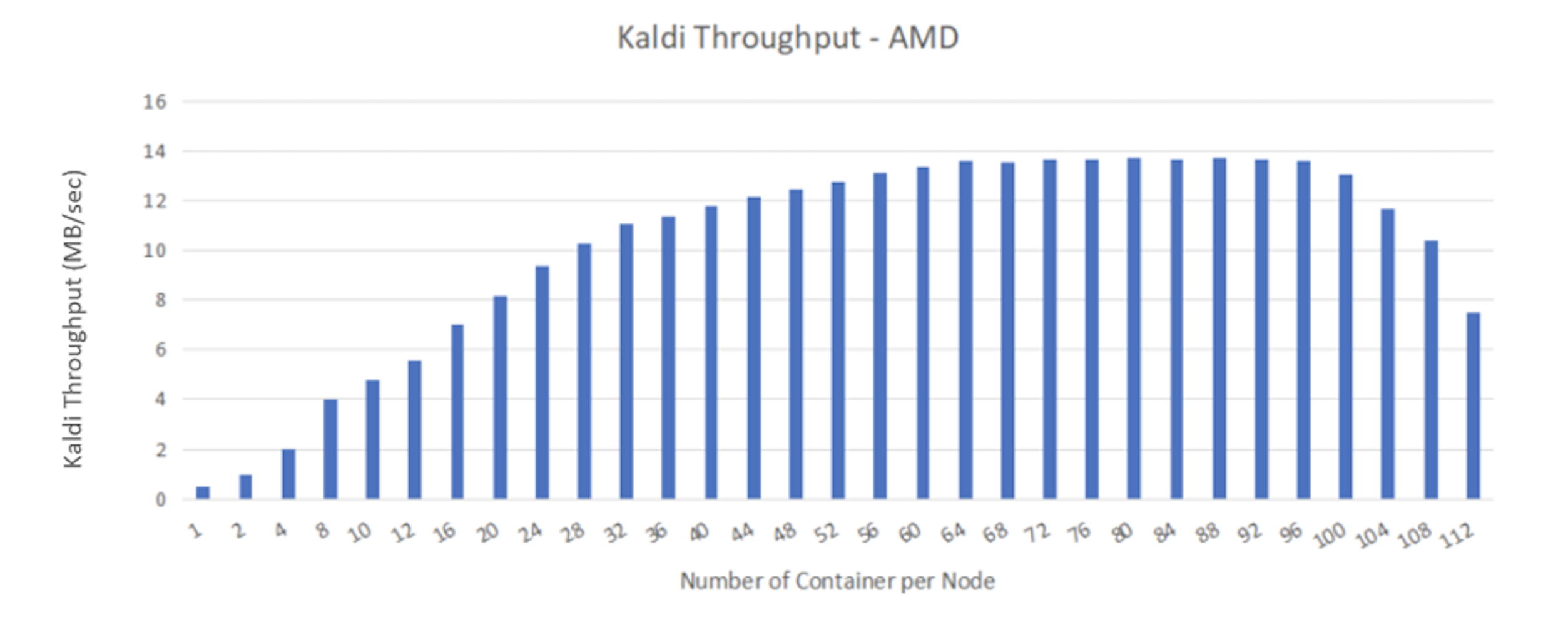}
\caption{Kaldi container performance vs. number of containers in AMD platform. Three levels of speech recognition completion time are observed. 1-to-64 containers as level 1, 64-to-196 containers as level 2, and 96-to-112 containers as level 3.}
\label{ContainerNumber_VS_KaldiPerformance_AMD}
\end{center}
\end{figure}

Figure~\ref{ContainerNumber_VS_KaldiPerformance_AMD} shows the Kaldi performance for different numbers of containers on an AMD server. Each container memory allocation is set to 1.3 GB. Three levels of performance were observed: at the first level, the system throughput increases (0.5 MB/sec to 13.8 MB/sec) along with the number of containers increases (0 to 64 containers). Then at the second level (64 to 96 containers), the system throughput flattens out and stabilizes at 13.8 MB/sec. Lastly, at the third level (96 to 112 containers), we observe that the system throughput quickly diminishes (13.8 MB/sec to 7.7 MB/sec) as the number of containers deployed increases. 

\subsubsection{System Analysis}
\label{sec-page-cache-amd}
\begin{figure*}[t]. 
\begin{center}
\includegraphics[width=1.4\columnwidth]{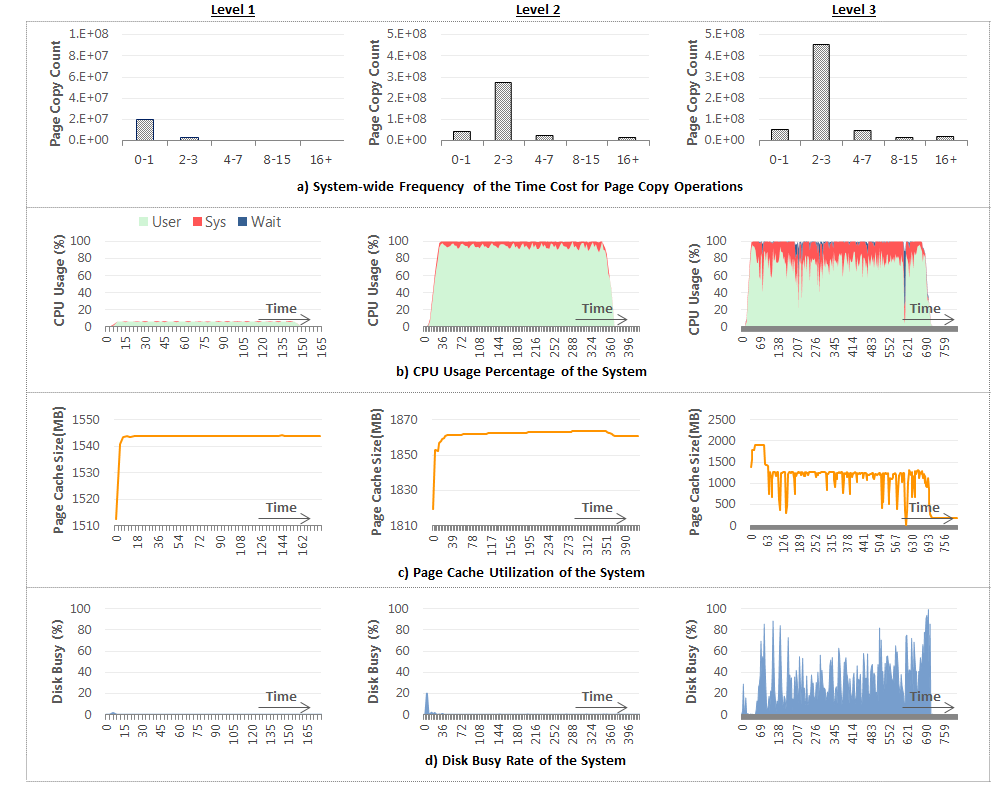}
\caption{System analysis plots with Kaldi containers on an AMD server: (a) the frequency of the time cost for page copy operations. (b) the CPU usage of the system (c) the page cache utilization over the time. (d) the disk busy rate of the system over time.}
\label{ContaineScaling_System_Analysis_AMD}
\end{center}
\end{figure*}

To understand the performance difference among three levels, for each Kaldi container deployment in Figure~\ref{ContaineScaling_System_Analysis_AMD} we analyze the corresponding time series of the ``under-the-hood" system resource usage of the whole AMD cloud server system. Since the data points at each level display a similar trend, we select one as representative (4 containers at level 1, 64 containers at level 2, and 104 containers at level 3) for demonstration purposes. 

Similar to the Intel servers, Figure~\ref{ContaineScaling_System_Analysis_AMD}(a) summarizes the frequency of the time cost for each page copy operation of the whole system. The x-axis represents the time cost per page copy operation in the microseconds range (i.e., 0-1 us), and the y-axis represents the total number of page copy operations in that range. During the speech recognition, each Kaldi container periodically loads the pre-trained language models. Such models are buffered in the virtual memory (``page cache" in Linux kernel) from the second time of access. So Kaldi containers can refer the data directly from the page cache without disk access taking place. However, all the containers in the system share the same page cache in the kernel, which leads to intensive data copy operations from the kernel space to the user space. The operational cost cannot be neglected, especially when the system is under a high workload. As Figure~\ref{ContaineScaling_System_Analysis_AMD}(a) and ~\ref{ContaineScaling_System_Analysis_AMD}(b) shows, each page copy operation takes around 1 microsecond to complete at four containers deployment case with a light CPU usage of the system. The time cost becomes worse when 64 containers are deployed on the server, in which most copy operations take 2 or 3 microseconds to finish, and the system CPU usage is already maximized. At 104 containers, most copy operations take 2 or 3 microseconds to finish, while more copy operations take 4 to 7 microseconds, or even 16+ microseconds to finish. The more containers deployed on the server, the intensive copy operations increase bus bandwidth pressure between the CPU and the subsystem, such as memory and I/O. At level 1, the pressure is relatively small as not too many containers are deployed. While at level 2 and level 3, the bus bandwidth pressures become difficult to handle in a full CPU utilization scenario, and therefore more and more copy operations require 2+ microseconds to finish or even 16+ microseconds at level 3. The accumulated time cost of each copy operation eventually generates the Kaldi performance overhead shown in Figure~\ref{ContainerNumber_VS_KaldiPerformance_AMD} along with the increased number of containers.  

Figure~\ref{ContaineScaling_System_Analysis_AMD}(c) and~\ref{ContaineScaling_System_Analysis_AMD}(d) show the system-wide page cache size and the busy disk rate over time. A steady and high page cache utilization, combined with a low disk busy rate, indicates all the deployed containers leverage the page cache well to load language models from RAM without disk access, as Figure~\ref{ContaineScaling_System_Analysis_AMD}(c) shows at levels 1 and 2. However, the page cache utilization grows and shrinks dynamically over time, as Figure~\ref{ContaineScaling_System_Analysis_AMD}(c) shows at level 3. This significant page cache fluctuation starts from 96 containers deployed on the server. As a consequence, all deployed containers have to interact with disk I/O at a slow rate. The more containers deployed on a server, the more contention would then happen at the high busy rate point, which leads to a significant Kaldi performance degradation (i.e., 1078 seconds completion time on average for 108 containers vs. 643 seconds for 104 containers). 

This significant page cache fluctuation is caused by the Linux cache eviction mechanism, which is triggered when the system is under severe memory pressure. Each Kaldi container memory limit is set to 1.3 GB; however, such memory limit setting only imposes the upper boundary of the container memory usage but not the memory reservation. At the container runtime later, the operating system allocates the container's requested memory on the fly. That said, if too many containers are running on a server, and if those container applications are memory intensive, the operating system kernel does not have enough free memory to allocate further requests. It has to reclaim the memory from the page cache and write the data back to the disk in order to reclaim memory. All containers running on a cloud server share the same page cache in the kernel; this reclamation thereby leads to system impact for all the other running containers and will force them to interact with the slow disk I/O to retrieve the evicted items. In the container scale case, disk access rather than page cache access is not desired, leading to the overall system slowdown. Moreover, all running containers also share the same buffer in the kernel for the disk I/O without any isolation nor any rate throttling mechanism per container cgroup. That said, the contention may happen in the disk I/O buffer as well and result in performance overhead: if a container/cgroup has a high disk I/O, that will interfere with all other containers running on the same server, which makes an even worse system impact for all other containers requiring disk I/O access as well. As a result, Kaldi container performance at level 3 will be severely degraded, as the tail part of Figure~\ref{ContainerNumber_VS_KaldiPerformance_AMD} shows. 

\subsubsection{Summary}
When deploying Kaldi at scale on an AMD-based cloud server, our results expose the non-negligible data copy operation overhead and the possible page cache eviction and its impact. Analysis of the utilization of CPU, page cache usage, and disk busy rate suggests the copy operation overhead is caused by the underlying bus bandwidth pressure between the CPU and subsystems such as memory and I/O. Although page cache is a common Linux kernel component to enhance the data access efficiency, the page cache eviction can be triggered under a system memory pressure case and will force all influenced running containers to directly interact with the disk. A secondary disk I/O contention can happen as all running containers share the same kernel buffer for disk I/O. From our observation, the page cache eviction can severely impact all the containers running on a cloud server, resulting in significant container performance degradation.

\section{Container Performance with Intel and AMD processors}
\label{sec:intel-vs-amd}
One interesting observation from the Kaldi experiments is the performance difference between the Intel and AMD platform. A comparison determines that the PCIe 4.0 technology adopted on AMD EPYC Rome chip is the main reason that makes container performance in the AMD platform outperforms the Intel platform. 

For both platforms, we re-use the same Kaldi 5 nodes topology as shown in the previous sections with the same CloudLab setup. In the topology, we still use the 1080 seconds audio file, with each file divided into 72 chunks. For both Intel and AMD platforms, we deploy four containers per Kaldi worker node, and for each container, we still set the memory limit to 1.3 GB. We then measure the Kaldi job completion time, along with the consumed system resources. 

\begin{figure}[htbp]   
\begin{center}
\includegraphics[width=0.95\columnwidth]{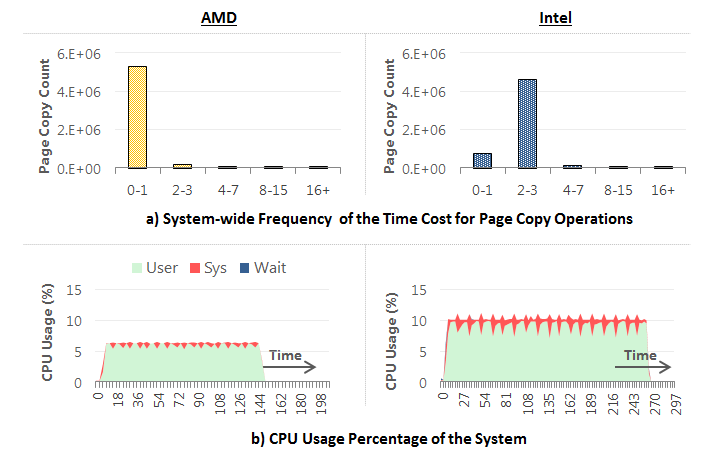}
\caption{Kaldi performance: AMD vs. Intel. (a) shows the frequency of time cost for each page copy operations for Intel/AMD platform, while (b) shows the system CPU usage for Intel/AMD platform.}
\label{AMD_vs_Intel}
\end{center}
\end{figure}

After conducting experiments, we see that Kaldi containers take around 145 seconds using AMD EPYC Rome cloud platform, while it takes around 255 seconds for the same setup using the Intel Xeon cloud platform. We analyze the time series of resource usage from both platforms, as shown in Figure~\ref{AMD_vs_Intel}. As Figure~\ref{AMD_vs_Intel}(a) shows, the most frequency of time cost for each page copy operation of the AMD platform is within 1 microsecond, but each page copy operation takes around 2 or 3 microseconds for the Intel platform. On CPU usage wise, for the same workload, AMD uses approximately 6.3\% of the system CPU resource while Intel utilizes around 9.7\%.

\begin{table}[htbp] 
\begin{center}
\caption{AMD Hardware vs. Intel Hardware}
\small
\begin{tabular}{|c|c|c|}
    \hline
    \textbf{\textit{Specification}} & \textbf{\textit{AMD Platform}} & \textbf{\textit{Intel Platform}} \\
    \hline
    Processor type & AMD 7452 & 2 Intel Xeon E5-2660 v2\\
    \hline
    CPU speed & 2.35GHz & 2.20 GHz\\
    \hline
    Max vCPU cores & 64 & 40\\
    \hline
    Memory size & 128GB & 256GB \\
    \hline
    Max DDR4 speed & 3200MT/s & 1600MT/s \\
    \hline
    PCIe Gen & 4.0 & 3.0 \\
    \hline
\end{tabular}
\label{Hardware_AMD_vs_Intel}
\end{center}
\end{table}

As table~\ref{Hardware_AMD_vs_Intel} listed, the AMD platform has a total of 64 vCPU cores while the Intel platform has 40 vCPU cores. With that being said, for the same Kaldi workload, both the Intel platform and AMD platform would use approximately the same amount of CPU resources as 64 * 6.3\% is approximately equal to 40 * 9.7\%. Since Kaldi requires loading a large size of a language model for incoming audio chunks, the workload pressure requires a great deal of bandwidth between the CPU and other subsystems such as memory and I/O, which essentially stresses the shared bus interconnecting them. Also, as table~\ref{Hardware_AMD_vs_Intel} listed, both Intel and AMD platform have almost the same CPU speed, so it cannot be said that one will significantly outperform the other. Indeed, PCIe 4.0 in AMD processors has twice the bandwidth compared to PCIe 3.0 in Intel processors. If workload stresses both Intel and AMD, the more data can be sent to the CPU, the faster each page copy operation job will be completed in the CPU. In our case, the 1 microseconds difference of page copy operation between two platforms will be accumulated and eventually gives the whole container performance difference in seconds level (144 seconds AMD vs. 255 seconds Intel). Going further, in the large-scale scenario, the more Kaldi workloads we have, the bigger container performance difference we will observe between the Intel platform and AMD platform.

\section{Deployment Guidance: Deploying Containerized Data-intensive Applications At Scale}
\label{sec:Dep}
To help cloud operators to make proper scale-up choices, we derive a list of practical guidelines below from experiment results. The goal is to help a broad range of similar containerized data-intensive application deployment, even though their actual data processing and data movement patterns have differences. 

\textbf{CPU Resources}: For data transferring applications, since they are often computation-intensive, cloud operators can leverage modern multi-core processor architecture on the server to distribute the computation task to achieve better performance. However, this wisdom also comes with a potential underlying cache level contention. Hence, it will be better to allocate CPU resources in a cache-aware manner, in which cloud operators should isolate containers based on L1/L2 cache to avoid the possibility of cache contentions. In addition, to bring efficiency, such core allocation distribution can potentially saturate the underlying memory data bus. To fully utilize the multi-core feature to scale up, cloud operators should place container deployment to a server with adequate data bus bandwidth, such as PCIe 4.0.

\textbf{Memory Resources}: For data processing applications, our results show that increasing the number of containers can generally increase overall application performance. However, it is important for cloud operators to be aware that the underlying page cache eviction may severely slow down the running container performance. This can happen when deploying more containers than the server can support in memory. Since there are various operating systems on the cloud server, it then requires the cloud operator to characterize the page cache effects before the container deployment. Secondly, the underlying memory data bus bandwidth can be another scale-up throttling factor. By comparing the container performance between the Intel platform and AMD platform, we found there exists a big performance difference, and essentially the difference is due to the memory data bus bandwidth (PCIe 3.0 on the Intel platform and PCIe 4.0 on the AMD platform).

\textbf{Characterize ``under-the-hood" system activities}: Since data-intensive applications oftentimes ingest, process, and send out a large amount of data, also given the container's weak isolation model fact, it is a good chance that the scale of containerized data-intensive applications would pressure the underlying system at multiple different places (as paper experiments studied in previous sections). Hence, the characterization of those ``under-the-hood" system activities would be a valuable approach for cloud operators to gain meaningful system insights. To have a deeper understanding and characterize those ``under-the-hood" system activities (e.g., page cache effect, page cache eviction effect, CPU operation utilization, etc.), cloud operator can leverage some OS kernel-level observing and monitoring tool such as eBPF to collect, measure, visualize and characterize those events and their container performance impact, in which Linux system itself not able to achieve such global view of its running system.

\section{Conclusion And Future Work}
\label{sec:con}
Although containers have become more attractive for many cloud applications' deployments, cloud operators raise more concerns for ``data-intensive" applications as they heavily consume all kinds of server resources. To deploy these types of applications at scale using containers, the efficiency of server resources utilization, resource contention, and container placement strategy optimization are still challenging tasks that are yet to be resolved. To address these challenges, this paper starts with a performance factor survey that studies multi-core Intel and AMD cloud systems, the architecture overview of containers and data-intensive applications at scale to closely inspect the relationship among those performance-relevant factors. Then, this paper studies two containerized data-intensive applications by running Docker container experiments on CloudLab servers with resources such as modern Intel and AMD multi-core processors, PCIe 3.0/4.0 data bus, and other different types of RAM, disk, and network interfaces. Our experimental results lead to a number of key observations of container resource contention behaviors that can affect the data-intensive application performance. Then, to help cloud operators make proper container scaling decisions, a set of clear guidelines are derived for the deployment of similar data-intensive applications.

Today, there are many different types of resources existing in the cloud computing environment (e.g., CPU, memory, I/O, etc.), while there are more types of resources existing in different cloud infrastructures. For example, in a fog/edge cloud infrastructure environment, resources like GPUs, FPGAs, and TPUs are still lack characterized and studied. This paper provides only a subset of the conditions and resources in a modern cloud data center environment, while we still need to access some other subsets resources in the cloud, or even edge cloud environment, and study their container performance impact to generate a fully holistic picture of the system. For example, how do the network I/O resource parameters assure that the given containerized data-intensive application runs with optimal performance and cost-efficiency? Based on the holistic picture we learned, we can then propose an automated and more general deployment solution for containerized cloud applications to better help cloud operators make proper scale-up choices. In such a solution, QoS parameters and constraints can be automatically imposed in scheduling and provisioning policies to further help cloud operators scale a containerized application deployment.

\section{Acknowledgement}
This material is based upon work sponsored by the National Science Foundation under grants No. 1743363, 2027208, and 1935966. Any opinions, findings, and conclusions or recommendations expressed in this material are those of the authors and do not necessarily reflect the views of the National Science Foundation.

\bibliography{References}

\end{document}